\documentclass[%
 superscriptaddress,
 reprint,
 amsmath,
 amssymb,
 prl,
 showpacs,
 twocolumn,
 floatfix,
 nofootinbib,
]{revtex4-1}

\usepackage{graphicx}
\usepackage{dcolumn}
\usepackage{bm}
\usepackage{hyperref}
\usepackage{color}
\usepackage{braket}
\usepackage{url}
\usepackage{xcolor}
\usepackage{soul}
\usepackage{lineno}

\begin{document}

\title{Engineering van der Waals heterostructures for dispersion-selective meV-scale quantum sensing}

\author{Elizabeth A. Peterson} \affiliation{Theoretical Division, Los Alamos National Laboratory, Los Alamos, NM 87545, USA}


\begin{abstract}
Quantum sensing of meV-scale scattering and absorption of impinging particles with electrons in solid state detectors is a challenging technological advancement with the potential to enable breakthroughs in quantum information applications and studies of fundamental physics. However, a key obstacle for current sensing schemes is the difficulty in distinguishing the signals from particles of interest and from intrinsic excitations, like phonons or magnons. Here we propose a technique to selectively detect impinging particles based not only on their imparted energy, but specifically by their dispersion relations. By harnessing interfacial orbital hybridization in van der Waals heterostructures of Dirac materials, interlayer charge transfer may be promoted only for pre-selected impinging particles of interest. Using first-principles density functional theory (DFT) calculations of heterostructures of the layered Dirac materials ZrTe$_{5}$ and HfTe$_{5}$, we examine the effects of strain and layer number for successfully tuning orbital hybridization in their electronic structure. We demonstrate a proof-or-principle feasibility study for using Dirac materials to construct “dispersion filters” to be leveraged for next-generation meV-scale quantum sensors.
\end{abstract}

\maketitle

\section{\label{sec:intro}Introduction}

Many next-generation technologies, spanning quantum key distribution, biomedical imaging, and detection of cosmological particles like light dark matter depend upon reliable sensing of meV-scale phenomena [\cite{Hadfield2009, Kajihara2013, Takemoto2015, Hong2017, Echternach2018, Hochberg2018, Griffin2021B, Dastidir2022}. Narrow band gap solid-state detectors that leverage electron scattering or absorption of impinging particles have the potential to be excellent platforms for development of tunable meV-scale quantum sensors. Electrons, as fermions, are not subject to the obstacles associated with boson-mediated meV-scale energy thresholds; specifically, electron excitations bypass timescale challenges associated with excited phonons rapidly decaying into lower energy phonons with a continuum of possible energies due to the linear dispersion of acoustic phonons. An excited electron in the conduction band of a narrow gap semiconductor has fewer allowed states to decay to. A material can be selected to have a band gap at the energy scale corresponding to interactions between electrons and an impinging particle of interest, as has been demonstrated in prior studies ~\cite{Essig2016, Hochberg2018, Emken2019, DAMIC2019, Griffin2020, Geilhufe2020, Trickle2020A, EDELWEISS2020, Griffin2021B, Coskuner2021, Hochberg2021, Knapen2021A, Kahn2022, Gu2022, CDEX2022, Trickle2023, DAMIC-M2023, Zema2024}. When this type of particle impinges on the detector material electrons are excited from the valence band to the conduction band producing a signal that is read out. Solid-state detector platforms can also be doped or strained to further refine their band gaps to the required energies. The detector physics community has actively developed a suite of tools for reading out electronic excitation signals of rare events in solid-state detector platforms via charge amplification schemes ~\cite{SuperCDMS2018, DAMIC2019, SENSEI2019, SENSEI2020, SuperCDMS2020, SPLENDOR2025}. 

A key challenge in quantum sensing at the meV-scale, however, is the broad diversity of excitations that could produce false positive signals. Intrinsic excitations (e.g. phonons and magnons) and extrinsically-induced excitations (e.g. infrared photons or cosmological particles) are all capable of inducing meV-scale excitations in a solid-state detector. Erroneous signals from intrinsic excitations are generically stymied by performing experiments at extreme cryogenic temperatures. Background extrinsic excitations can be suppressed by performing the detection experiment in extreme isolation, such as at the bottom of a deep mine where ambient electromagnetic waves cannot penetrate deep enough into the Earth's crust to induce false signals. This strategy has been employed widely in the cosmological physics field, including for detection of particles of dark matter or neutrinos ~\cite{SENSEI2019, DAMIC2019, EDELWEISS2020, LZ2025}. However, this approach to removing background signals imposes inherent limitations on how many experiments can be conducted and how often.

For quantum sensing of particles in more terrestrial setting, there is a need for detection schemes that can straightforwardly distinguish between the desired particles and background sources of electronic excitation. In this work we outline a scheme to selectively detect particular impinging particles by engineering materials platforms that can detect both the energy \textit{and the momentum} transferred from the impinging particle to the excited electron. This scheme leverages orbital hybridization across heterostructures of van der Waals (vdW) materials. With knowledge of the dispersion relations of the impinging particle of interest, a quantum sensor can be engineered that selectively detects particles that impart the correct combination of energy and momentum. Here we offer a proof-of-principle example vdW heterostructure of layered Dirac materials that can be tuned to act as such a "dispersion filter" for quantum sensing.

\section{\label{sec:theory}Theoretical Framework}

\begin{figure}
   \includegraphics[width=1.0\linewidth]{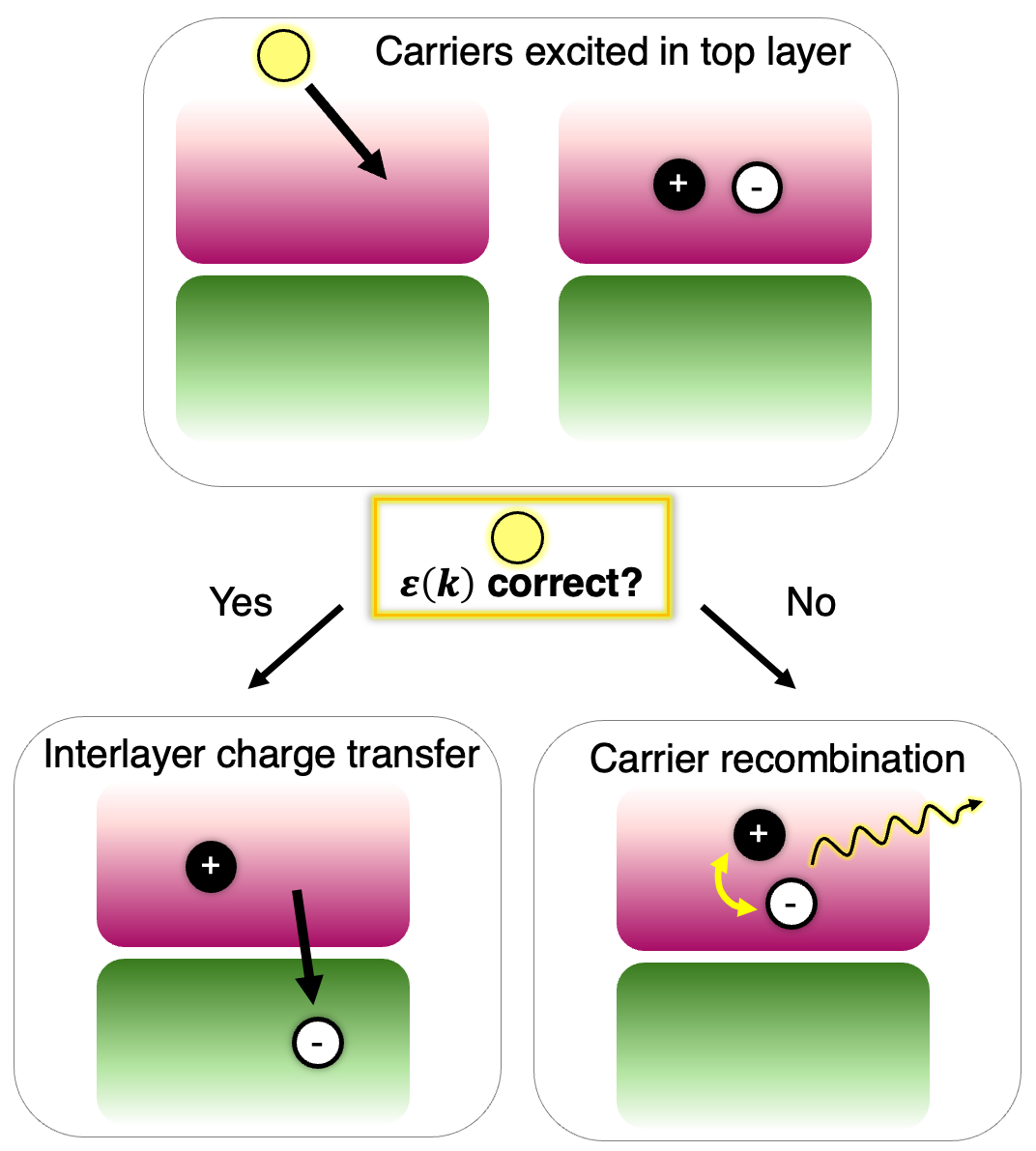}

\caption{\label{fig:real_space}Simplified Real-space schematic of detector setup. An impinging particle produces excited electrons and holes in the top layer. An excited electron rapidly transfers to the bottom layer only if the impinging particle had the correct dispersion relations. Otherwise carriers remain in the top layer and recombine.}
\end{figure}

The proposed detector architecture is built on a vdW heterostructure of two layered materials that each have distinct band gaps. The particle of interest impinges on the first layer producing excited electrons and holes. If and only if that impinging particle has the correct dispersion relations (allowed energy and momentum transfer), an excited electron (or hole) will transfer to the second layer (Figure \ref{fig:real_space}). The presence of the excited carrier in the second layer will signal that the desired particle of interest has been detected. Read out can be accomplished by charge amplification schemes already in use in the detector community which can resolve single-digit numbers of excited carriers ~\cite{SuperCDMS2018, DAMIC2019, SENSEI2019, SENSEI2020, SuperCDMS2020, SPLENDOR2025}.

For interlayer charge transfer to occur in multilayer vdW materials, carriers must tunnel through the interlayer vacuum region. Generically, charge transfer across a vacuum region should occur much more slowly than bulk or intralayer transport.

However, as has been demonstrated in multilayers of transition metal dichalcogenides (TMDs), ultrafast charge transfer can be facilitated by interlayer orbital overlap. An illustration of how this manifests in the bulk TMD MoS$_{2}$ is illustrated in the Supplementary Information (Figure S1). A representative example of this phenomenon occuring in a heterobilayer TMD, constructed of two distinct TMDs, was demonstrated by Sood, et al. ~\cite{Sood2023}. In heterobilayer WS$_{2}$ and WSe$_{2}$ there is a band offset between the direct gaps of each monolayer with the valence band maximum (VBM) and conduction band minimum (CBM) of WSe$_{2}$ shifted to higher energy than those of WS$_{2}$ (see Supplementary Information, Figure S2). The band gap of WSe$_{2}$ is smaller than that of WS$_{2}$ enabling selective optical excitation of WSe$_{2}$. When WSe$_{2}$ is optically excited, the excited electrons and holes initially exist only in the WSe$_{2}$ layer. The CBM of WS$_{2}$ is at lower energy than that of WSe$_{2}$, but in order for an electron to decay to the WS$_{2}$ conduction band state interlayer charge transfer through the vacuum layer must occur. In principle this would be expected to occur slowly, however the electronic structure of heterostructures are greater than the sum of their parts. The electron in the CBM of WSe$_{2}$ can rapidly decay to the CBM of WS$_{2}$ by passing first through a hybridized state in the $\Lambda$ valley whose orbital character spans both layers \cite{Sood2023}. An illustration of this process can be found in the Supplementary Information (Figure S3). The broad implication is that hybridized conduction states are a key tool for facilitating rapid interlayer charge transfer.

\begin{figure*}
   \includegraphics[width=0.9\linewidth]{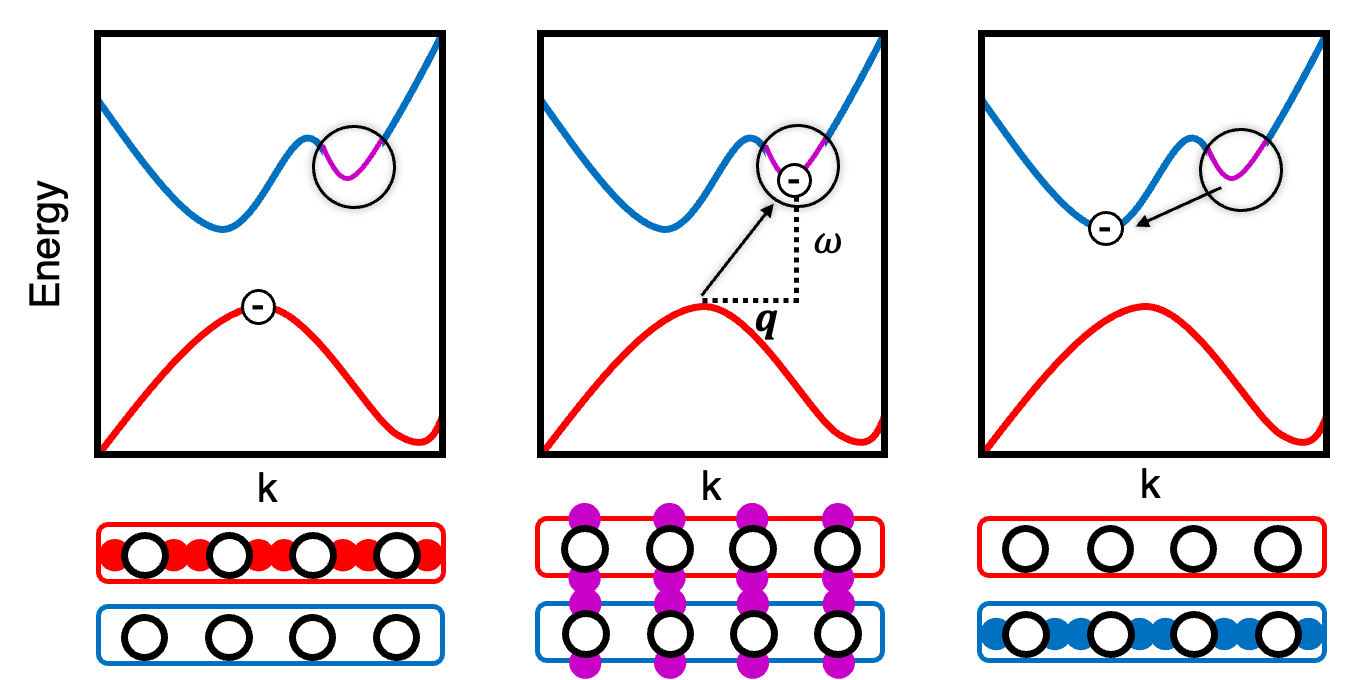}

\caption{\label{fig:k_space}Reciprocal space (top panels) and real space (bottom panels) schematics of proposed detection scheme. A valence electron at the VBM in layer 1 (top layer in real space panels) is excited to a hybridized state in the conduction band with orbital character spanning both layers. The excited electron then decays into the CBM and occupies an orbital that is entirely in layer 2 (bottom layer in real space panels). This process can only occur if the electron was excited by an impinging particle with the correct dispersion relations (combination of transferred energy ($\omega$) and transferred momentum ($\mathbf{q}$) (center reciprocal space panel)}
\end{figure*}

The scheme proposed in this work reverse-engineers this concept. The electronic structure of the detector heterostructure (Figure \ref{fig:k_space}) is such that the orbital character of the valence band maximum (VBM) is entirely restricted to the first layer and the orbital character of the conduction band minimum (CBM) is entirely restricted to the second layer. There is a valley in the conduction band, of slightly higher energy, that is of hybridized orbital character, with its wavefunction spanning both layers. The particle that one seeks to detect has appropriate dispersion relations to impart the correct combination of energy ($\omega$) and momentum ($\mathbf{q}$) to excite an electron from the VBM in layer 1 into the hybridized state (circled region of conduction band in Fig. \ref{fig:k_space}). Once there, because the wavefunction of the hybridized state that the excited electron occupies spans both layers, it readily decays to the CBM of layer 2. The presence of excess carriers in layer 2 signals that a particle with the correct dispersion relations has been detected. Read out of this signal would be achieved using charge amplification schemes, such as those used in rare event detection schemes used in dark matter searches ~\cite{SuperCDMS2018, DAMIC2019, SENSEI2019, SENSEI2020, SuperCDMS2020, SPLENDOR2025}.

\begin{figure*}
   \includegraphics[width=0.8\linewidth]{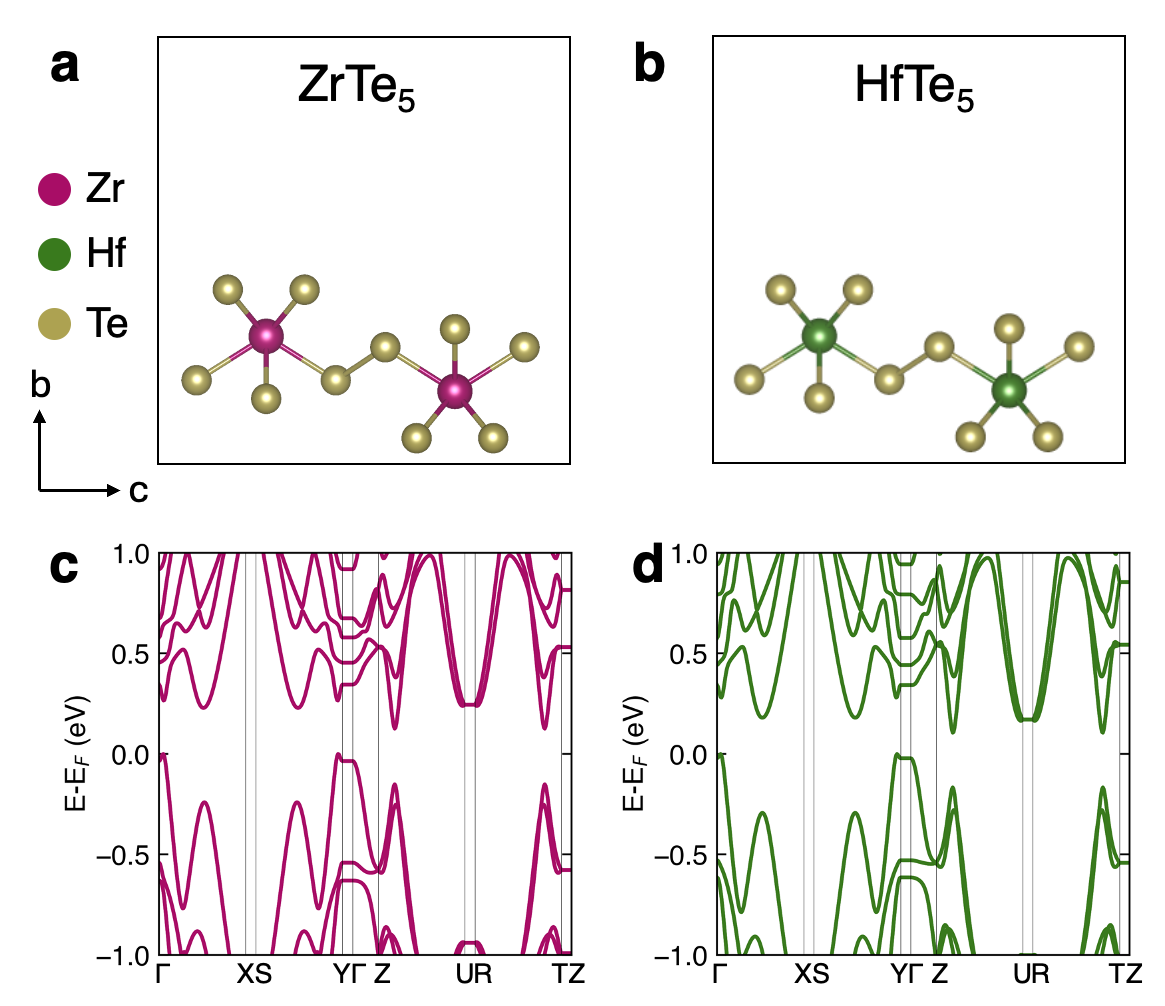}

\caption{\label{fig:monolayers}Crystal structures of (a) monolayer ZrTe$_{5}$ and (b) monolayer HfTe$_{5}$ and their corresponding band structures (c) and (d), respectively.}
\end{figure*}

The band gaps of the widely studied TMDs in the 2H structure are generally on the order of $\sim$ eV, which is too large to be sensitive to meV-scale impinging particles. To target meV-scale detectors, vdW materials with smaller band gaps are required. In this work we target the isostructural layered Dirac materials ZrTe$_{5}$ and HfTe$_{5}$. Monolayer versions of these materials are displayed in Figure \ref{fig:monolayers}(a)(b). These materials have been broadly studied in the bulk form. They exhibit a topological phase transition from a weak topological insulator, to a topological semimetal, to a strong topological insulator, which has been measured to be highly tunable, especially via strain ~\cite{Fan2017, Zhang2017, Mutch2019, Tajkov2022, Liu2024, Jo2024, Peterson2025}. Crystals as grown, in their unstrained geometry, are generally weak topological insulators with band gaps on the order of a couple dozen meV ~\cite{Zhang2017, Mutch2019, Liu2024, Jo2024, Peterson2025}. Their tunability and small band gaps render them a useful starting point for exploring meV-scale dispersion filter vdW heterostructure detectors.

\section{\label{sec:calc_details}Calculation Details}

First-principles calculations are performed using density functional theory (DFT) with a plane-wave basis and projector augmented wave (PAW) pseudopotentials ~\cite{Kresse1999} as implemented in the Vienna \textit{ab initio} simulation package (VASP) ~\cite{Kresse1996a,Kresse1996b}. Calculations are performed in the generalized gradient approximation (GGA) as implemented by Perdew, Burke, and Ernzerhof (PBE) ~\cite{Perdew1996} with additional vdW dispersion forces approximately accounted for via the Grimme-D3 method ~\cite{Grimme2010}. The crystal structures of ZrTe$_{5}$, HfTe$_{5}$, and their heterostructures are relaxed using a 500 eV energy cutoff and $18\,\times1\,\times6$ $\Gamma$-centered k-mesh until forces are converged to $<$ 1 meV/\AA. To prevent interactions between periodic images of the monolayers and heterostructures along the out-of-plane \textit{b-}axis, at least 20 \AA of vacuum is included in the out-of-plane direction. The calculated in-plane lattice parameters for monolayer ZrTe$_{5}$ are \textit{a} = 4.03 \AA and \textit{c} = 13.71 \AA, in good agreement with  experimental values (\textit{a} = 3.99 \AA , \textit{c} = 13.72 \AA )~\cite{Fjellvag1986}. The calculated lattice parameters for monolayer HfTe$_{5}$ are \textit{a} = 4.01 \AA and \textit{c} = 13.67 \AA, in reasonably good agreement with  experimental values (\textit{a} = 3.97 \AA,  \textit{c} = 13.73 \AA)~\cite{Fjellvag1986}. The calculated lattice parameters of the heterobilayer of ZrTe$_{5}$ and HfTe$_{5}$ are \textit{a} = 4.02 \AA, \textit{b} = 12.72 \AA , and \textit{c} = 13.66 \AA. This corresponds to effective strains of $-0.2\%$ and $+0.2\%$ along the in-plane \textit{a}-axis and $-0.4\%$ and $\sim 0.0\%$ along the in-plane \textit{c}-axis for ZrTe$_{5}$ and HfTe$_{5}$, respectively. Biaxial strain is applied by expanding and fixing the in-plane (\textit{a-} and \textit{c-}axis) lattice parameters then allowing the internal coordinates of the heterostructure to relax.  Both the geometry relaxations and band structures are calculated with spin-orbit coupling (SOC).

\section{\label{sec:results}Results}

A series of electronic structure calculations are performed on the representative isostructural layered Dirac materials ZrTe$_{5}$ and HfTe$_{5}$ (Figure \ref{fig:monolayers}(a,b)). These materials are well-known to have highly tunable electronic and topological properties, in particular under applied strain ~\cite{Fan2017, Zhang2017, Shahi2018, Monserrat2019, Mutch2019, Tajkov2022, Peterson2025, Jo2024, Liu2024}.

The DFT band structures of monolayer ZrTe$_{5}$ and HfTe$_{5}$ (Figure \ref{fig:monolayers}(c,d)) have calculated band gaps of 126 and 106 meV respectively. In previous work by Peterson, et al. ~\cite{Peterson2025}, the band structures of bulk ZrTe$_{5}$ and HfTe$_{5}$ were calculated using the same DFT methodology as employed in this work, yielding band gaps of 27 and 24 meV respectively. The order of magnitude increase in band gap moving from bulk crystals to monolayers indicates layer number is another powerful tuning knob for engineering the electronic structure of these materials, consistent with results observed for other layered vdW materials ~\cite{Mak2010, Rudenko2014, Kim2015, Gusakova2017}.

\begin{figure*}
    \centering
      \includegraphics[width=1.0\linewidth]{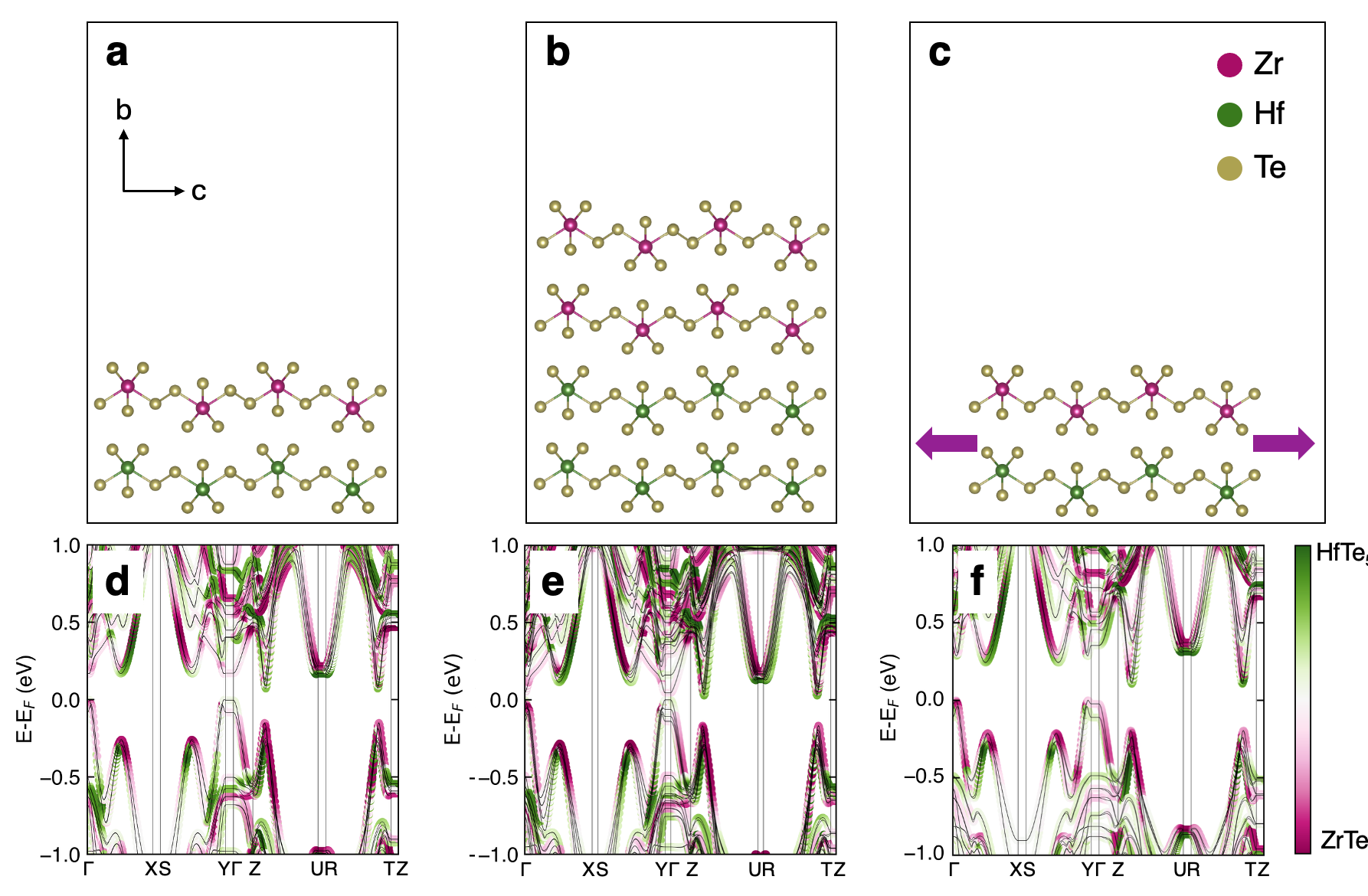}
    \caption{Crystal structures and electronic structures of various ZrTe$_{5}$-HfTe$_{5}$ heterostructures: (a,d) monolayer ZrTe$_{5}$ with monolayer HfTe$_{5}$, (b,e) bilayer ZrTe$_{5}$ with bilayer HfTe$_{5}$, and (c,f) monolayer ZrTe$_{5}$ with monolayer HfTe$_{5}$ under 3\% tensile strain. The band structures are colored according to orbital projections onto each layer with purple indicating wavefunctions primarily confined to ZrTe$_{5}$ and green indicating wavefunctions primarily confined to HfTe$_{5}$. Lighter colored and white areas have hybridized wavefunctions with contributions from orbitals that span both layers.}
    \label{fig:heterostructures_bs}
\end{figure*}

The band structure of heterobilayer ZrTe$_{5}$ and HfTe$_{5}$ with layer-resolved orbital projections (Figure \ref{fig:heterostructures_bs}(a) is plotted in Figure \ref{fig:heterostructures_bs}(d). The band gap of this heterostructure is 72 meV, of intermediate value between the bulk and monolayer band gaps.  Unfortunately for the purposes of the proposed detector scheme, the VBM of this heterobilayer is highly hybridized with orbital character spanning both layers rendering it undesirable for dispersion selective interlayer charge transfer.

The band structure of a heterostructure of bilayer ZrTe$_{5}$ and bilayer HfTe$_{5}$ (Figure \ref{fig:heterostructures_bs}(b) exhibits a very substantial reduction in band gap to 27 meV. This result is consistent with the known sensitivity of vdW materials to number of layers and demonstrates that this property can be used as another tuning knob to control the excitation energy threshold. However, the persistence of hybridization in the VBM again renders this heterostructure undesirable for dispersion selective interlayer charge transfer.

The final tuning knob explored is epitaxial strain. The band structures of the heterobilayer under both compressive and tensile biaxial strain up to $\pm$ 3\% were calculated and the orbital character of the bands was analyzed (see Supplementary Information Figures S4 and S5).  At 3\% tensile biaxial strain the band structure of the heterobilayer of monolayer ZrTe$_{5}$ and monolayer HfTe$_{5}$ (Figure \ref{fig:heterostructures_bs}(c) exhibits a substantial change in the orbital character of the band extrema near the gap. The VBM is at the $\mathrm{Y}$ high symmetry point and consists primarily of orbitals localized on ZrTe$_{5}$. The CBM located between the $\mathrm{Z}$ and $\mathrm{U}$ high-symmetry points consists primarily of orbitals localized on HfTe$_{5}$. At slightly higher energy than the CBM are hybridized states near the $\mathrm{Y}$ and $\mathrm{\Gamma}$ high symmetry points. This combination of orbital character at the VBM, CBM, and slightly higher energy points along the CB matches the combination in the theoretical framework outlined in Figure \ref{fig:k_space}. These results illustrate the potential feasibilty of engineering narrow band gap vdW heterostructures for meV-scale "dispersion filter" particle detectors.

\section{\label{sec:summary}Summary}

In this work we outline a novel framework for meV-scale particle detection. The key innovation is that our detection scheme is designed to circumvent false positive results by restricting positive signals to interactions between electrons and impinging particles with a pre-determined dispersion relation, meaning the combinations of energy and momentum that the impinging particle's dispersion rules allow. The key to the development of this "dispersion filter" scheme is interlayer orbital hybridization in vdW layered heterostructures. We provide a proof-of-principle demonstration using first-principles calculations showing that the electronic structure of vdW heterostructures of layered Dirac materials can be tuned to have the appropriate orbital hybridization to promote dispersion-selective interlayer charge transfer. Here we use DFT to study the electronic structure of heterobilayers of ZrTe$_{5}$ and HfTe$_{5}$ and show that under 3\% tensile biaxial strain the VBM adopts ZrTe$_{5}$ orbital character, the CBM adopts HfTe$_{5}$ orbital character, and the region of the CB slightly above the CBM has highly hybridized orbital character, spanning both layers.

These results open a pathway to selectively detect particles of interest without requiring the extreme methods that state-of-the-art particle detectors require for background elimination. In principle, a dispersion-selective detector built from narrow band gap materials could be used under cryogenic tabletop conditions, substantially reducing the cost associated with background signal removal. This type of detector could open a new field of detector physics because it possesses the power to easily differentiate between impinging particles with the same energy but different momenta. This proof-of-principle result has the potential to open a new pathway to tackle the broad array of applications that are dependent upon meV-scale quantum sensing.

\section{\label{sec:ackn}Acknowledgements}
This work was supported by the U.S. DOE NNSA under Contract No. 89233218CNA000001. The work was funded by the LANL LDRD Program through the Institute for Materials Science (IMS) Rapid Response program and project number 20220135DR. This work was supported in part by the Center for Integrated Nanotechnologies, a DOE Office of Science user facility, in partnership with the LANL Institutional Computing Program for computational resources.  Additional computations were performed at the National Energy Research Scientific Computing Center (NERSC), a U.S. Department of Energy Office of Science User Facility located at Lawrence Berkeley National Laboratory, operated under Contract No. DE-AC02-05CH11231 using NERSC award ERCAP0020494.


\providecommand{\noopsort}[1]{}\providecommand{\singleletter}[1]{#1}%

\clearpage

\begin{Large}
\begin{center}
\begin{widetext}
Supplementary Information: Engineering van der Waals heterostructures for dispersion-selective meV-scale quantum sensing
\end{widetext}
\end{center}
\end{Large}


\setcounter{figure}{0}

\renewcommand{\thefigure}{S\arabic{figure}}

\begin{figure*}
   \includegraphics[width=0.9\linewidth]{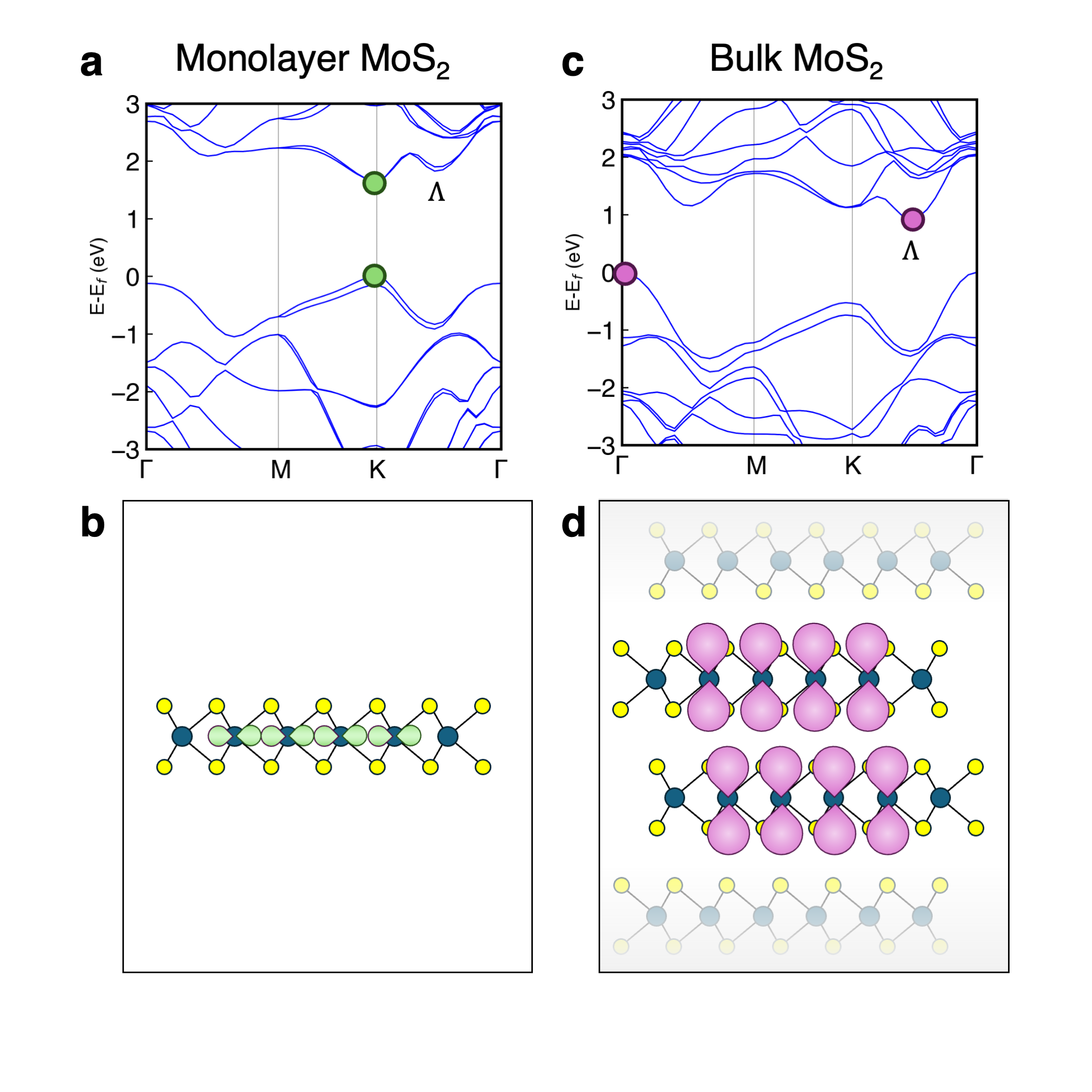}

\caption{\label{fig:mono_bulk_bs}The band structures of monolayer and bulk MoS$_{2}$. (a) In the band structure for monolayer MoS$_{2}$ the valence band maximum (VBM) and conduction band minimum (CBM) are both located at the high symmetry $\mathrm({K}$ point. (b) The orbital character of the VBM is entirely in-plane (orbitals are illustrated in green). (c) In the band structure for bulk MoS$_{2}$ the VBM is located at the high symmetry $\mathrm({\Gamma}$ point and CBM is located at a minimum in the $\mathrm({\Lambda}$ valley. (d) The orbital character of VBM and CBM in bulk MoS$_{2}$ includes strong out-of-plane contributions (orbitals are illustrated in purple). This leads to interlayer hybridization.} 
\end{figure*}

\begin{figure*}
   \includegraphics[width=0.7\linewidth]{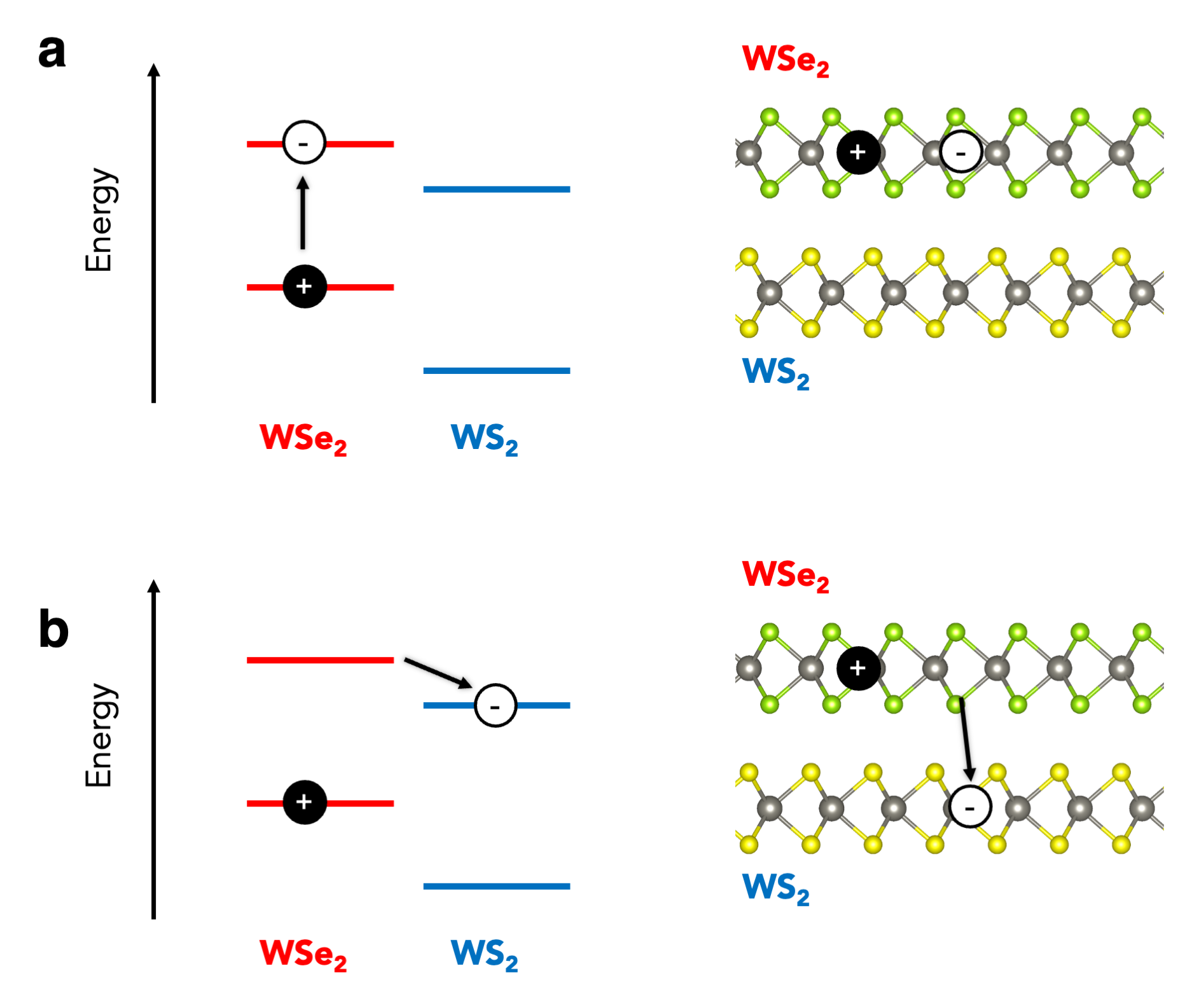}

\caption{\label{fig:het_band_offset}Band offset between WSe$_{2}$ and WS$_{2}$ in a heterobilayer. When (a) an electron in WSe$_{2}$ is optically excited to the CBM of WSe$_{2}$, it rapidly tunnels through the van der Waals gap between the two layers to (b) the CBM of WS$_{2}$. This phenomenon was measured in ~\cite{Sood2023}.} 
\end{figure*}

\begin{figure*}
   \includegraphics[width=1.0\linewidth]{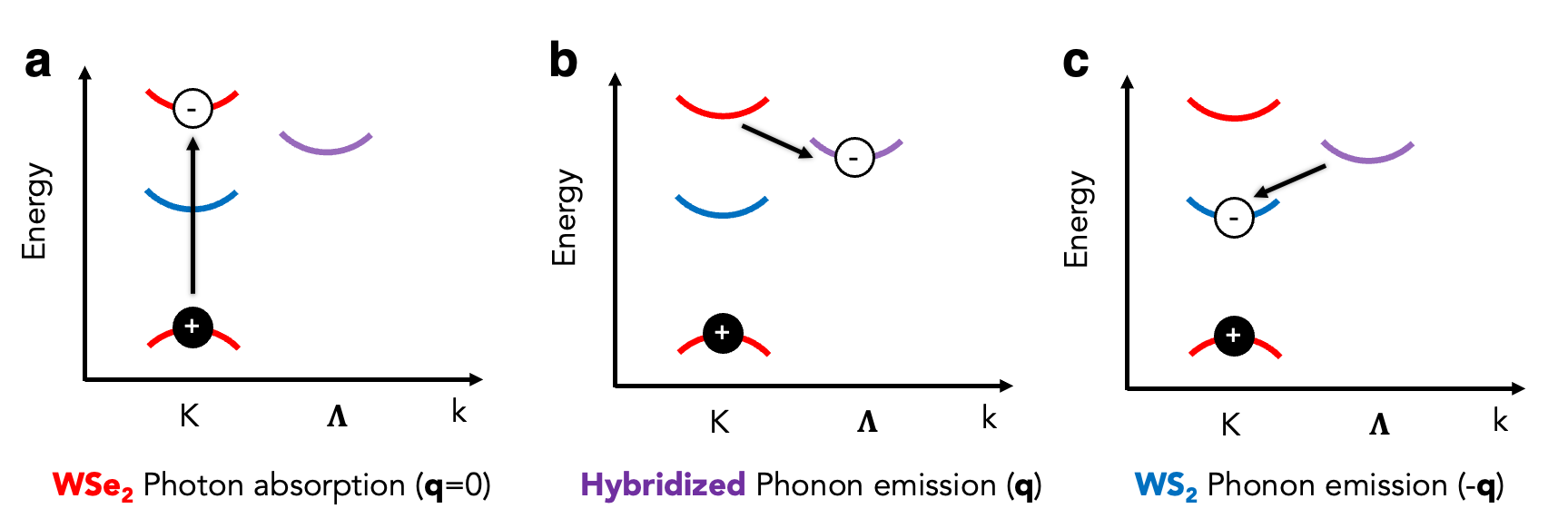}

\caption{\label{fig:het_bs}The mechanism by which an excited electron in the CBM of WSe$_{2}$ rapidly tunnels to the CBM of WS$_{2}$. (a) The electron is optically excited from the VBM to the CBM of WSe$_{2}$. (b) The electron in the CBM of WSe$_{2}$ releases a phonon and decays to the strongly hybridized minimum of the $\mathrm{\Lambda}$ valley, which has orbital character spanning both layers. (c) The electron in the $\mathrm{\Lambda}$ valley, which is already resonantly present in WS$_{2}$, releases another phonon and decays to the CBM of WS$_{2}$. This phenomenon has been described in detail in ~\cite{Sood2023}.} 
\end{figure*}

\begin{figure*}
   \includegraphics[width=1.0\linewidth]{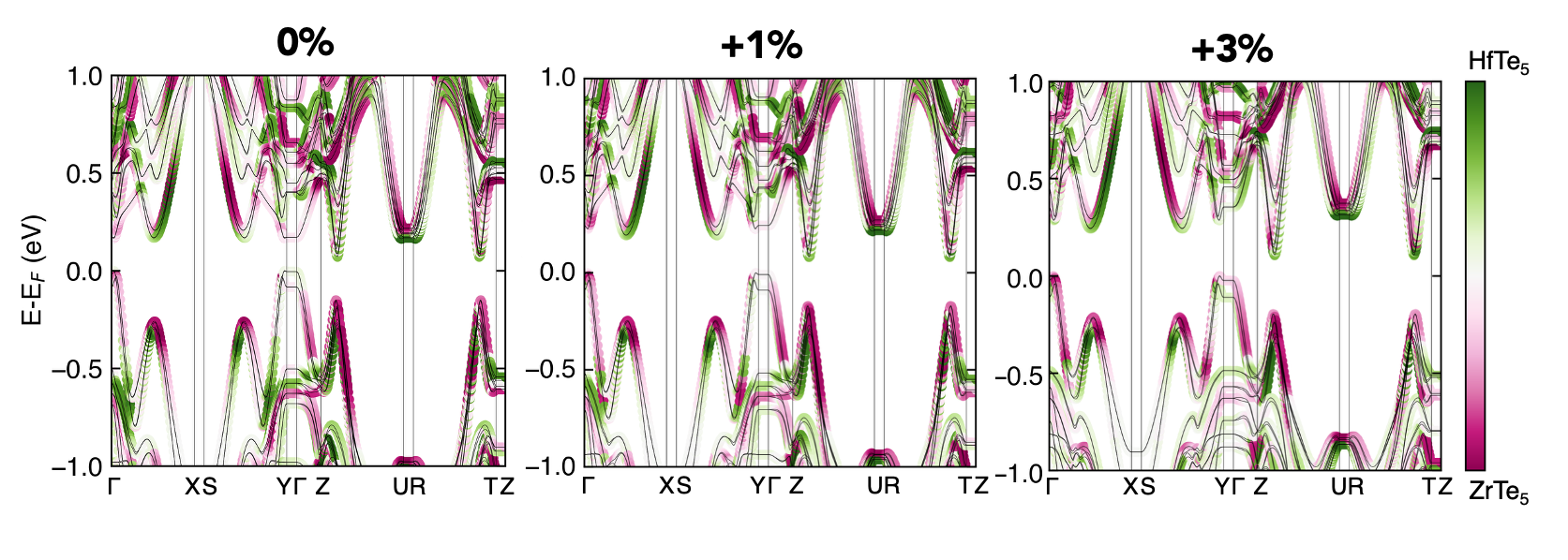}

\caption{\label{fig:tensile_bs}Band structures of heterobilayers of monolayer ZrTe$_{5}$ with monolayer HfTe$_{5}$ under (a) 0\% (b) 1\% and (c) 3\% biaxial tensile strain. The band structures are colored according to orbital projections onto each layer with purple indicating wavefunctions primarily confined to ZrTe$_{5}$ and green indicating wavefunctions primarily confined to HfTe$_{5}$. Lighter colored and white areas have hybridized wavefunctions with contributions from orbitals that span both layers.}
\end{figure*}

\begin{figure*}
   \includegraphics[width=1.0\linewidth]{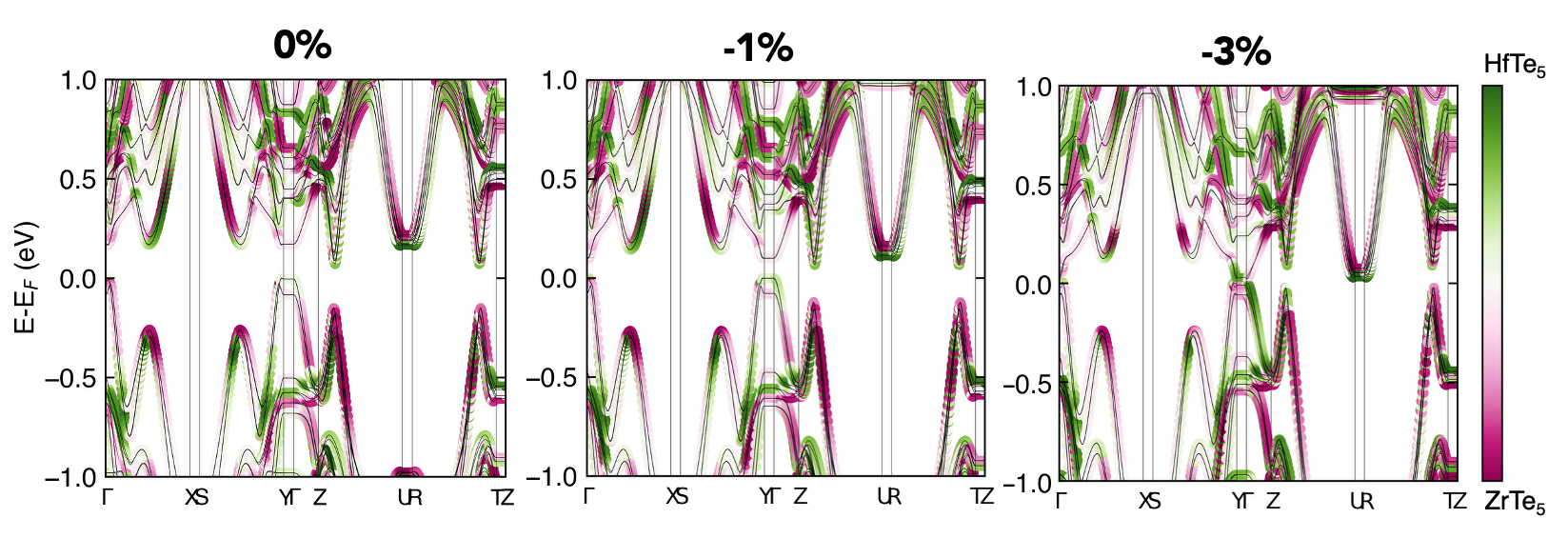}

\caption{\label{fig:compressive_bs}Band structures of heterobilayers of monolayer ZrTe$_{5}$ with monolayer HfTe$_{5}$ under (a) 0\% (b) 1\% and (c) 3\% biaxial compressive strain. The band structures are colored according to orbital projections onto each layer with purple indicating wavefunctions primarily confined to ZrTe$_{5}$ and green indicating wavefunctions primarily confined to HfTe$_{5}$. Lighter colored and white areas have hybridized wavefunctions with contributions from orbitals that span both layers.}
\end{figure*}

\end{document}